\documentclass{SciPost}

\hypersetup{
    colorlinks,
    linkcolor={red!50!black},
    citecolor={blue!50!black},
    urlcolor={blue!80!black}
}

\usepackage[bitstream-charter]{mathdesign}
\usepackage{physics}

\DeclareSymbolFont{usualmathcal}{OMS}{cmsy}{m}{n}
\DeclareSymbolFontAlphabet{\mathcal}{usualmathcal}

\fancypagestyle{SPstyle}{
\fancyhf{}
\lhead{\colorbox{scipostblue}{\bf \color{white} ~SciPost Physics }}
\rhead{{\bf \color{scipostdeepblue} ~Submission }}

\fancyfoot[C]{\textbf{\thepage}}
}

\begin{document}

\newcommand{\half}{\frac{1}{2}}
\newcommand{\quarter}{\frac{1}{4}}

\newcommand{\Rdot}{\dot{R}}
\newcommand{\udot}{\dot{u}}
\newcommand{\Pdot}{\dot{P}}
\newcommand{\pdot}{\dot{p}}
\newcommand{\Sdot}{\dot{S}}
\newcommand{\sdot}{\dot{s}}
\newcommand{\Pidot}{\dot{\Pi}}
\newcommand{\pidot}{\dot{\pi}}
\newcommand{\AR}[1][]{A^{(R)}_{#1}}
\newcommand{\AS}[1][]{A^{(S)}_{#1}}
\newcommand{\Au}[1][]{A^{(u)}_{#1}}
\newcommand{\As}[1][]{A^{(s)}_{#1}}

\newcommand{\Kuu}{{K^{uu}}}
\newcommand{\Ksu}{{K^{su}}}
\newcommand{\Kus}{{K^{us}}}
\newcommand{\Kss}{{K^{ss}}}

\newcommand{\Duu}[1][]{{D^{uu}_{#1}}}
\newcommand{\Dsu}[1][]{{D^{su}_{#1}}}
\newcommand{\Dus}[1][]{{D^{us}_{#1}}}
\newcommand{\Dss}[1][]{{D^{ss}_{#1}}}

\newcommand{\Dsutilde}{{\tilde D^{su}}}
\newcommand{\Dustilde}{{\tilde D^{us}}}
\newcommand{\Dsstilde}{{\tilde D^{ss}}}

\newcommand{\Guu}[1][]{{G^{uu}_{#1}}}
\newcommand{\Gsu}[1][]{{G^{su}_{#1}}}
\newcommand{\Gus}[1][]{{G^{us}_{#1}}}
\newcommand{\Gss}[1][]{{G^{ss}_{#1}}}

\newcommand{\Gsutilde}{{\tilde G^{su}}}
\newcommand{\Gustilde}{{\tilde G^{us}}}
\newcommand{\Gsstilde}{{\tilde G^{ss}}}

\newcommand{\omegaqv}{\omega_{q\nu}}
\newcommand{\xqv}{x_{q\nu}}
\newcommand{\epsqv}{\varepsilon_{q\nu}}
\newcommand{\epsqvk}{\varepsilon_{q\nu,\kappa}}
\newcommand{\epsqvtilde}{\tilde\varepsilon_{q\nu}}
\newcommand{\epsqvktilde}{\tilde\varepsilon_{q\nu,\kappa}}
\newcommand{\muqv}{\mu_{q\nu}}
\newcommand{\muqvk}{\mu_{q\nu,\kappa}}
\newcommand{\muqvtilde}{\tilde\mu_{q\nu}}
\newcommand{\sigmaqv}{\sigma_{q\nu}}

\newcommand{\lqvzero}{l_{q\nu}^0}

\pagestyle{SPstyle}

\begin{center}{\Large \textbf{\color{scipostdeepblue}{
Theory of phonon-magnon hybridization and angular momentum in CrI$_3$ and CrBr$_3$
}}}\end{center}

\begin{center}\textbf{
Maxime Mignolet\textsuperscript{1,2,3$\star$},
Miquel Royo\textsuperscript{4},
Massimiliano Stengel\textsuperscript{4,5} and
Matthieu J. Verstraete\textsuperscript{1,3,6$\dagger$}
}\end{center}

\begin{center}
{\bf 1} Nanomat Group, Q-Mat, Université de Liège, B-4000 Liège, Belgium
\\
{\bf 2} Fonds de la Recherche Scientifique, B-1000 Bruxelles, Belgium
\\
{\bf 3} European Theoretical Spectroscopy Facility
\\
{\bf 4} Institut de Ciència de Materials de Barcelona (ICMAB-CSIC), Carrer dels Til.lers, 08193 Cerdanyola del Vallès, Spain
\\
{\bf 5} ICREA-Institució Catalana de Recerca i Estudis Avançats, 08010 Barcelona, Spain
\\
{\bf 6} ITP, Department of Physics, University of Utrecht, 3508TA Utrecht, The Netherlands
\\[\baselineskip]
$\star$ \href{mailto:M.Mignolet@uliege.be}{\small M.Mignolet@uliege.be}\,,\quad
$\dagger$ \href{mailto:m.j.verstraete@uu.nl}{\small m.j.verstraete@uu.nl}
\end{center}

\section*{\color{scipostdeepblue}{Abstract}}
\textbf{\boldmath{%
In magnetic materials angular momentum can be mediated by different carriers, including electrons, magnons, and phonons. The magnons can interact with circularly polarized phonons which are close in energy, provided specific symmetry conditions are met. If the interaction is strong enough the phonons and magnons shift in frequency and start to mix to form hybrid magneto-elastic quasi-particles. In this paper, we develop a constrained Hamiltonian framework which incorporates hybrid stiffness matrices and Berry curvatures.
We quantify the degree of hybridization between phonons and magnons (up to 8\% in CrI$_3$ and 25\% in CrBr$_3$) using a decomposition of the total energy, which is a generalization of the norm decomposition for atomic contributions to phonons used in the literature.
We also explore how the total angular momentum is conserved but shared between the phononic and magnonic subsystems upon hybridization.
}}

\vspace{\baselineskip}

\noindent\textcolor{white!90!black}{%
\fbox{\parbox{0.975\linewidth}{%
\textcolor{white!40!black}{\begin{tabular}{lr}%
  \begin{minipage}{0.6\textwidth}%
    {\small Copyright attribution to authors. \newline
    This work is a submission to SciPost Physics. \newline
    License information to appear upon publication. \newline
    Publication information to appear upon publication.}
  \end{minipage} & \begin{minipage}{0.4\textwidth}
    {\small Received Date \newline Accepted Date \newline Published Date}%
  \end{minipage}
\end{tabular}}
}}
}


\vspace{10pt}
\noindent\rule{\textwidth}{1pt}
\tableofcontents
\noindent\rule{\textwidth}{1pt}
\vspace{10pt}


\section{Introduction}

Phonon angular momentum (PAM) has attracted broad interest since its prediction more than a decade ago \cite{zhang14}. It has come to light that phonon angular momentum plays a central role in many phenomena, such as the Einstein-de Haas effect \cite{zhang14}, ultrafast demagnetization processes \cite{tauchert22} and the thermal Hall effect \cite{strohm05,inyushkin07,grissonnanche20}.
Novel phenomena centered around PAM have also been unveiled, e.g. the phonon magnetic moment and the phonon Zeeman effect \cite{schaack76,schaack77,juraschek17,chaudhary24}, the phonon diode effect \cite{chen22b} and the PAM Hall effect \cite{park20,Lopez26}.
Phonon angular momentum has been investigated as a mechanism for possible dark matter detectors \cite{romao23b}, as well as for spin current injection in spintronics application \cite{funato24}.
PAM has been observed numerous times, both indirectly through phonon-assisted intervalley electron-hole recombination in WSe$_2$ \cite{zhu18}, and directly through ultrafast electron diffraction in ferromagnetic Ni \cite{tauchert22}, by inelastic X-ray scattering in WC \cite{cai21} and chiral quartz \cite{ueda23}, and by measurement of the torque induced by generation of PAM by a temperature gradient in chiral Te \cite{zhang25a}. 

Phonon angular momentum requires symmetry breaking to be well-defined and non-degenerate \cite{coh23,juraschek25}.
The breaking can involve the spatial inversion symmetry which can be found in non-centrosymmetric materials. It can be induced, for example, by a static external electric field \cite{hamada20,sonntag21,moseni22a}, or driven by a time-dependent external electric field \cite{geilhufe21,basini24}. 
The symmetry breaking can also concern time reversal symmetry, if the material itself is magnetic \cite{coh23,bonini23}, subject to an external magnetic field \cite{schaack76,schaack77}, or to a rigid-body rotation \cite{wang15}. 

Phonon angular momentum can interact and couple with other angular momenta, such as spins in magnetic materials. This interaction between PAM and spins can be modeled phenomenologically via a Raman spin-lattice coupling \cite{sheng06,kagan08}. Other modeling approaches use either an explicit magneto-elastic energy term, or a magnetic (Heisenberg) exchange coupling that depends on the interatomic distances \cite{Flebus17,Streib19,Weissenhofer25}.

The ab initio study of phonon angular momentum in magnetic materials goes through the inclusion of the molecular Berry curvature \cite{coh23,saparov22}, possibly including the dynamic of atomic magnetic moments \cite{bonini23, ren24a} into the equations of motion for lattice dynamics. The molecular Berry curvature is an adiabatic correction resulting from the contribution of the electronic cloud to the nuclear kinetic energy and canonical momentum \cite{mead79, mead92}. In the equations of motion, it manifests itself as a velocity-dependent force analogous to a (non-local) magnetic field. A similar approach is possible through dynamical spin-lattice coupling~\cite{Fransson17,Royo26}.

In this paper, we develop a rigorous unified quantum theory of coupled phonon-magnon dynamics and hybridization. We treat the ionic and (atomic) spin degrees of freedom on the same footing. We reformulate the preexisting equations of motion to highlight particle-hole symmetry and the link with the total energy. We derive a novel orthonormalization scheme for the eigenmodes and establish a measure of the phonon-magnon hybridization, based on the partitioning of the total energy, which is one of the main results of this paper. Finally, we examine the exchange of angular momentum between magnons and phonons upon hybridization, which had not been quantified in the literature up to now.

The layout of the paper is as follows. Section \ref{sec:theory} details the theoretical developments made, which is subdivided in four parts:
the derivation of the Lagrangian and Hamiltonian dynamics alongside the equations of motion,
the second quantization of the framework, determining the total energy and quantifying the hybridization,
and a discussion of the link between the current theory and Raman spin-lattice interaction.
Section \ref{sec:results} presents applications of the theory to first-principles computations through the study of phonon-magnons in bulk CrI$_3$ and CrBr$_3$.
Lastly in section \ref{sec:conclusion}, we conclude and discuss possible future developments and applications.

\section{Theory}\label{sec:theory}

\subsection{Lagrangian and Hamiltonian mechanics}

In this section, we will derive the equations of motion for coupled spin-nuclei dynamics.
We denote $R=\{R_i\}$ the positions of the atoms. We define the mass-weighted displacement $u_i$ of atom $i$ from its equilibrium position $R_{i0}$ as $u_i = \sqrt{M_i}(R_i-R_{i0})$.
We note $s=\{s_i\}$ the deviations of the atomic spins from their equilibrium values $S_{i0}$, such that $s_i=S_{i}-S_{i0}$. We will consider a ferromagnetic material for which we have $S_{i0}=S_0 \; \forall i$. We use the notation $\udot$, $\sdot$ to denote their time derivatives. $u$, $\udot$, $s$ and $\sdot$ will constitute our dynamical variables. We consider that $u$ and $s$ evolve sufficiently slowly such that the electronic system remains in its ground state and we can write the wavefunction as a parametric function of $u$ and $s$, which we denote $\Psi(r;u,s)$. 
The system's Lagrangian is given by
\begin{equation}
    \mathcal{L}(u,\udot,s,\sdot,t) = \frac{\udot^2}{2} + \mel{\Psi}{i\hbar\frac{d}{dt} - H_e}{\Psi}
\end{equation}
where $H_e$ is the electronic Hamiltonian and includes the electron-electron, electron-ion and ion-ion Coulomb interactions, and the electronic kinetic energy. Considering that the dynamical variables are functions of time and using the chain rule, the total time derivative can be written as:
\begin{equation}
    \frac{d}{dt} = \partial_t + \udot_i \frac{\partial}{\partial u_i} + \sdot_i \frac{\partial}{\partial s_i}.
\end{equation}
This yields the following Lagrangian in $u$ and $s$:
\begin{equation}
    \mathcal{L} = \frac{\udot^2}{2} + i\hbar \udot_i \expval{\partial_{u_i}} + i\hbar \sdot_i \expval{\partial_{s_i}} - V
\end{equation}
where $V=\expval{H_e}$ and where we used the short-hand notation $\partial / \partial x = \partial_x$. We can perform a Legendre transform with respect to $u$, $\udot$ and $s$, $\sdot$. The conjugate momenta are given by
\begin{align}
    p_i & = \frac{\partial \mathcal{L}}{\partial \udot_i} = \udot_i +  \Au[i],\\
    \pi_i & = \frac{\partial \mathcal{L}}{\partial \sdot_i} = \As[i],
\end{align}
where we defined the atomic and spin Berry connections $\Au[i] = i\hbar \expval{\partial_{u_i}}$, $\As[i] = i\hbar \expval{\partial_{s_i}}$ \footnote{To lighten the equations, the factor $\hbar$ has been absorbed in the usual definition of the Berry connections.}.
The first relation can be inverted to obtain an expression of the velocity $\udot$ as a function of the position $u$ and momentum $p$. However the same cannot be done for $\sdot$ with the second relation. Instead, we formulate the constraint $\phi_i=\pi_i-\As[i] \approx 0$. The "$\approx$" sign means that the constraint should only be imposed after the equations of motion have been obtained \cite{dirac01,anderson51,brown22a}.
The constrained Hamiltonian arising from the Legendre transform is written as:
\begin{align}
    & \mathcal{H}(u,p,s,\pi,t) = p_i \udot_i + \pi_i \sdot_i - \mathcal{L} + \lambda_i \phi_i\\
    \Leftrightarrow & \mathcal{H} = \frac{1}{2} \Big(p_i - \Au[i]\Big)^2 + V + \lambda_i \phi_i, \label{eq:constrained_Ham}
\end{align}
where $\lambda_i$ are Lagrange multipliers.
The appeal of the Hamiltonian formulation over the Lagrangian is that it offers a natural path to obtain an expression for the total energy of the system and a decomposition of the eigenmodes in phononic, magnonic and coupled contributions. This will be shown in Sec. \ref{sub:total_energy}.

Since we started from an effective adiabatic Lagrangian formulation, the current approach does not include the Born contribution to the total energy \cite{Born51,Moody89}. However the Born contribution does not break time reversal symmetry and is almost always neglected \cite{saparov22,bonini23,ren24a}. If needed, the Born contribution can be reincorporated while maintaining a Lagrangian starting point as detailed in Ref. \cite{Moody89}. 

From the above Hamiltonian we can derive the coupled equations of motion for the atomic spins and positions. Using Hamilton's equations, we obtain:
\begin{align}
    & \udot_i = \frac{\partial H}{\partial p_i} = p_i - \Au[i],\\
    & \pdot_i = - \frac{\partial H}{\partial u_i} = \left(p_j - \Au[j] \right) \partial_{u_i} \Au[j] - \partial_{u_i} V - \lambda_j \partial_{u_i} \phi_j,\\
    & \sdot_i = \frac{\partial H}{\partial \pi_i} = \lambda_j \partial_{\pi_i} \phi_j,\label{eq:lambda}\\
    & \pidot_i = - \frac{\partial H}{\partial s_i} = \left(p_j - \Au[j] \right) \partial_{s_i} \Au[j] - \partial_{s_i} V - \lambda_j \partial_{s_i} \phi_j.
\end{align}
We can now impose the constraint and use it to eliminate $\pi$. We can further use Eq. \eqref{eq:lambda} to eliminate $\lambda$. We obtain:
\begin{align}
    & \udot_i = p_i - \Au[i],\\
    & \pdot_i = \left(p_j - \Au[j] \right)\partial_{u_i} \Au[j] - \partial_{u_i} V + \sdot_j \partial_{u_i} \As[j],\\
    & \dot{A}^{(s)}_i = \left(p_j - \Au[j] \right)\partial_{s_i} \Au[j] - \partial_{s_i} V + \sdot_j \partial_{s_i} \As[j].
\end{align}
If we place ourselves near an equilibrium reference configuration, the potential energy $V$ and the Berry connections $\Au$ and $\As$ can be expanded as Taylor series in $u$ and $s$. The potential energy is expanded up to second order to retrieve the harmonic approximation. The Berry connections are expanded up to first order to obtain the Berry curvatures which are gauge-invariant:
\begin{align}
    V=\frac{1}{2} \begin{pmatrix} u & s \end{pmatrix}
      \begin{pmatrix} \Kuu & \Kus \\ \Ksu & \Kss \end{pmatrix}
      \begin{pmatrix} u \\ s \end{pmatrix}, \label{eq:harmonic_approx} \\
    \begin{pmatrix} \Au \\ \As \end{pmatrix} = - \frac{1}{2}
      \begin{pmatrix} \Guu & \Gus \\ \Gsu & \Gss \end{pmatrix}
      \begin{pmatrix} u \\ s \end{pmatrix}. \label{eq:berryCurvaturesDef}
\end{align}
For simplicity, we assumed a symmetric gauge for the Berry connections \cite{saparov22}. The matrices $\Guu$, $\Gss$, and $\Gus$ and $\Gsu$ are the molecular Berry curvature, the spin Berry curvature, and the mixed "spin-molecular" Berry curvatures respectively. Their general definition is given by
\begin{align}
    G^{\mu\nu} & = \partial_\mu A_\nu - \partial_\nu A_\mu \nonumber \\
    & = i\hbar \Big( \braket{\partial_\mu \psi}{\partial_\nu \psi} - \braket{\partial_\nu \psi}{\partial_\mu \psi} \Big)
\end{align}
From this definition, it can be seen that the Berry curvatures $\Guu$ and $\Gss$ are real antisymmetric: $\Guu = -(\Guu)^t$, $\Gss = -(\Gss)^t$. $\Gus$ and $\Gsu$ are real and the antisymmetric of one another: $\Gus = -  (\Gsu)^t$.
The equations of motion become:
\begin{subequations} \label{eq:EOM_Hamiltonian}
\begin{align}
    \udot_i & = \half \Guu[ij] u_j + p_i + \half \Gus[ij] s_j,\\
    \pdot_i - \half \Gus[ij] \sdot_j & = -\Duu[ij] u_j + \half \Guu[ij] p_j - \Dus[ij] s_j,\\
    \half \Gsu[ij] \udot_j + \Gss[ij] \sdot_j & = \Dsu[ij] u_j - \half \Gsu[ij] p_j + \Dss[ij] s_j,
\end{align}
\end{subequations}
where we introduced the following matrices:
\begin{align}
    \Duu & = \Kuu - \quarter \Guu\Guu,\\
    \Dus & = \Kus - \quarter \Guu\Gus,\\
    \Dss & = \Kss - \quarter \Gsu\Gus.
\end{align}
And we have $\left(\Duu\right)^t=\Duu$, $\left(\Dus\right)^t=\Dsu$, $\left(\Dss\right)^t=\Dss$.
The system of equations Eq. \eqref{eq:EOM_Hamiltonian} can be recast in matrix form by multiplying the second line by minus one and permuting it with the first:
\begin{align}
    \mathcal{M} \dot{X} = \mathcal{D} X,\quad
    X =
    \begin{pmatrix}
        u \\ p \\ s
    \end{pmatrix} \label{eq:EOM_real_short}
\end{align}
with the matrices:
\begin{align}
    \mathcal{M} & =
    \begin{pmatrix}
        0 & -1 & \half \Gus \\
        1 & 0 & 0 \\
        \half \Gsu & 0 & \Gss
    \end{pmatrix}, \label{eq:M_matrix_real}\\
    \mathcal{D} & =
    \begin{pmatrix}
        \Duu & - \half \Guu & \Dus\\
        \half \Guu & 1 & \half \Gus\\
        \Dsu & - \half \Gsu & \Dss
    \end{pmatrix}. \label{eq:D_matrix_real}
\end{align}
Since the $D^{xx}$ and $G^{xx}$ matrices are real symmetric and antisymmetric, the matrices $\mathcal{M}$ and $\mathcal{D}$ are real antisymmetric and symmetric, respectively.
Eq. \eqref{eq:EOM_real_short} admits the general solution $X(t)=X e^{\Omega t}$. This yields the generalized eigenvalue problem
\begin{align}
    \Omega_\lambda \mathcal{M} X_\lambda = \mathcal{D} X_\lambda.\label{eq:gen_eig_probl_classical}
\end{align}
Since $\mathcal{D}$ is real symmetric and $\mathcal{M}$ is real antisymmetric, the eigenvalues are purely imaginary and come in complex conjugated pairs $i\omega$, $-i\omega$. 
There are $3 \times N_{at}+N_{s}$ modes with positive frequency, and the same number with a negative frequency.
The pairs bear the same physical meaning and correspond to the same physical solution.
Furthermore, the corresponding eigenvectors $X_\lambda$ also come in complex conjugated pairs. They can be $\mathcal{M}$-orthonormalized through $X_\lambda^t \mathcal{M} X_{\lambda'} = \pm i \delta_{\lambda\lambda'}$. The choice for the prefactor will be made clear in Sec. \ref{sub:total_energy}.

From Eq. \eqref{eq:EOM_real_short}, \eqref{eq:M_matrix_real} and \eqref{eq:D_matrix_real}, we can recover the equations for the bare phonons by discarding the spin parts and the molecular Berry curvature (which embodies adiabatic corrections):
\begin{align}
    \begin{pmatrix}
        0 & -1 \\ 1 & 0
    \end{pmatrix}
    \begin{pmatrix}
        \udot \\ \pdot
    \end{pmatrix}
    & =
    \begin{pmatrix}
        \Duu & 0 \\ 0 & 1
    \end{pmatrix}
    \begin{pmatrix}
        u \\ p
    \end{pmatrix}\\[8pt]
    \Leftrightarrow &
    \ddot u = - \Duu u. \label{eq:bare_phonon}
\end{align}
This equation is the standard equation for phonons\cite{Born98}. For spins, by discarding the ionic degrees of freedom, one obtains
\begin{align}
    \Gss \sdot = \Kss s. \label{eq:bare_magnon}
\end{align}
This is the equation of motion derived by Niu and Kleinman \cite{niu98}, and has been shown to simplify to the Landau-Lifshitz equation under the assumption of rigid spins \cite{niu99}.

\subsection{Second Quantization}\label{sub:second_quantization}

We proceed to second quantize the equations of motion in Eq. \eqref{eq:EOM_real_short}. First, we perform a Fourier transform to make use of the periodicity of the problem:
\begin{align}
    \hat x_{l,\kappa} = \frac{1}{\sqrt{N}} \sum_q \hat x_{q,\kappa} e^{iqR_{l,\kappa}},\quad x=\{u,p,s\} \label{eq:FT_real2recip}
\end{align}
The matrices transform accordingly. 
No mass factor appears here as $u,p$ are already mass-scaled.

We now introduce the Holstein Primakoff transformation for the spins \cite{holstein40,guerreiro71}. For this, we switch from a cartesian representation to a +/- ladder representation: 
\begin{align}
    \begin{pmatrix}
        s_{+} \\ s_{-}
    \end{pmatrix}
    = \frac{1}{\sqrt{2}}
    \begin{pmatrix}
        1 &  i\\
        1 & -i
    \end{pmatrix}
    \begin{pmatrix}
        s_{x} \\ s_{y}
    \end{pmatrix}
    = \mathcal{T}^{-1}
    \begin{pmatrix}
        s_{x} \\ s_{y}
    \end{pmatrix},
\end{align}
where the $q,\kappa$ indices have been omitted for clarity.
This transformation is unitary. The matrices pertaining to the spins transform as:
\begin{align}
    \Dsstilde = \mathcal{T}^{-1} \Dss \mathcal{T},\;\; \Dustilde = \Dus \mathcal{T}.
\end{align}
The Berry curvatures transform similarly. The pure phonon part of the hessian matrix ($\Duu$) and of the Berry curvature ($\Guu$) are left unchanged. This is a similarity transform of the $\mathcal{M}$ and $\mathcal{D}$ matrices. Therefore, it leaves the eigenvalues unchanged.

We can now introduce the Holstein-Primakoff transformation
\begin{subequations}
    \label{eq:HP}
    \begin{align}
    s_{i,+} & = \sqrt{S_0} \sqrt{1 - a_i^\dagger a_i/2S_0} \;a_i,\label{eq:HP1}\\
    s_{i,-} & = a_i^\dagger \;\sqrt{S_0}\sqrt{1 - a_i^\dagger a_i/2S_0},\label{eq:HP2}\\
    s_{i,z} & = S_{i,z} - S_{0} = - a_i^\dagger a_i,\label{eq:HP3}
    \end{align}
\end{subequations}
where $i=(q,\kappa)$. The Holstein-Primakoff operators obey the usual bosonic commutation relations: $[a_i,a_j^\dagger]=\delta_{ij}$, $[a_i,a_j]=[a_i^\dagger,a_j^\dagger]=0$. Relations \eqref{eq:HP1}, \eqref{eq:HP2} can be linearized within the large S limit \cite{auerbach94,utermohlen20}:
\begin{subequations}
    \label{eq:HPlin}
    \begin{align}
    s_{i+} = \sqrt{S_0} \;a_i,\label{eq:HPlin1}\\
    s_{i-} = \sqrt{S_0} \;a_i^\dagger.\label{eq:HPlin2}
    \end{align}
\end{subequations}

In order to diagonalize the Hamiltonian and solve the equations of motion, we use a Bogoliubov transform \cite{anderson51,holz72,erickson91,udvardi03}:
\begin{align}
    u_{q \kappa} & = \frac{1}{\sqrt{2}} ( \varepsilon_{q\kappa} b_q + \varepsilon_{-q\kappa}^* b_{-q}^\dagger ),\label{eq:Bogoliubov_displ}\\
    p_{q \kappa} & = \frac{1}{\sqrt{2}} ( \mu_{q\kappa} b_q + \mu_{-q\kappa}^* b_{-q}^\dagger ),\label{eq:Bogoliubov_momentum}\\
    a_{q \kappa} & = \frac{1}{\sqrt{2}} ( \sigma_{q\kappa,+} b_q + \sigma_{-q\kappa,-}^* b_{-q}^\dagger ),
\end{align}
where we made use of $u_{q \kappa}^* = u_{-q \kappa}$, $p_{q \kappa}^* = p_{-q \kappa}$, which stems from $u_{l \kappa}, p_{l \kappa}$ being real, to reduce the number of free parameters. The eigenvectors have components along $u$, $p$, and $s$ which are $\varepsilon$, $\mu$ and $\sigma$. We note that $\varepsilon_{-q\nu,\kappa}^*$ does not reduce to $\varepsilon_{q\nu,\kappa}$ in general since time reversal symmetry is broken. The bosonic creation/annihilation operators obey the standard commutation relations: $[b_i,b_j^\dagger]=\delta_{ij}$, $[b_i,b_j]=[b_i^\dagger,b_j^\dagger]=0$.
The semiclassical equations of motion \eqref{eq:EOM_real_short} now become:
\begin{align}
    -i\omega_{q\nu} \tilde{\mathcal{M}} \Xi_{q\nu} b_{q\nu} = \tilde{\mathcal{D}}\Xi_{q\nu} b_{q\nu},\quad
    \Xi_{q\nu} =
    \begin{pmatrix}
        \varepsilon_{q\nu} \\ \mu_{q\nu} \\ \sigma_{q\nu}
    \end{pmatrix} \label{eq:EOM_quantized_short}
\end{align}
with the new matrices:
\begin{align}
    \tilde{\mathcal{M}} & =
    \begin{pmatrix}
        0 & -1 & \half \Gustilde \\
        1 & 0 & 0 \\
        \half \Gsutilde & 0 & \Gsstilde
    \end{pmatrix},\\
    \tilde{\mathcal{D}} & =
    \begin{pmatrix}
        \Duu & - \half \Guu & \Dustilde\\
        \half \Guu & 1 & \half \Gustilde\\
        \Dsutilde & - \half \Gsutilde & \Dsstilde
    \end{pmatrix},
\end{align}
and where we used $\dot{b}_{q\nu} = -i\omega_{q\nu} b_{q\nu}$. As it was the case for Eq. \eqref{eq:gen_eig_probl_classical}, the eigenfrequencies obtained from Eq. \eqref{eq:EOM_quantized_short} come in pairs of positive and negative frequencies: $\omega_{q,-\nu} = -\omega_{q,\nu}$.
It is customary to assign positive frequency solutions to the annihilation operator $b_{q\nu}$ and negative frequency solutions to the creation operator $b_{-q\nu}^\dagger$ \cite{holz72}. In that sense, there are $3*N_{at}+N_{s}$ independent modes, although there is double the amount of solutions to Eq. \eqref{eq:EOM_quantized_short}.

Before pursuing with the next sub-section, it is noteworthy to mention that the factors $\sqrt{S_0}$ introduced in the Holstein-Primakoff transformation Eq.~\eqref{eq:HPlin} effectively bring a rescaling of the spin-related matrices as $\Dsstilde \mapsto S_0 \Dsstilde,\ \Dsutilde \mapsto \sqrt{S_0} \Dsutilde$, and similarly for the other matrices (see Ref. \cite{toth15}).
In the case of non-equivalent spins $S_{i0}\neq S_{j0}$, the rescaling would become: $\Dsstilde_{ij} \mapsto \sqrt{S_{i0}S_{j0}} \Dsstilde_{ij},\ \Dsutilde_{ij} \mapsto \sqrt{S_{i0}} \Dsutilde_{ij}$. This is in total analogy to the mass rescaling of the dynamical matrix in the standard theory of phonons \cite{Bruesch82,gonze97a}.

\subsection{Total energy and phonon-magnon hybridization} \label{sub:total_energy}

We now turn our attention towards the total energy of the system. We restart from the Hamiltonian from Eq. \eqref{eq:constrained_Ham}. By using the Harmonic approximation for the potential energy and the Taylor expansion of the Berry connections from Eq. \eqref{eq:harmonic_approx} and \eqref{eq:berryCurvaturesDef}, we obtain the following expression for the total energy:
\begin{align}
    E = \frac{1}{2} X^t \mathcal{D} X.
\end{align}
The constraint does not contribute to the total energy due to it being identically zero for solutions satisfying the equations of motion. Plugging in the solutions from Eq. \eqref{eq:gen_eig_probl_classical}, we obtain the expression of the total energy in real space in the semi-classical regime:
\begin{equation}
    E = \half \sum_\lambda \Omega_\lambda X_\lambda^H \mathcal{M} X_\lambda.
\end{equation}
The superscript $H$ denotes the complex conjugate transpose. Using the second quantization procedure from the preceding sub-section, the total energy becomes
\begin{align}
    E = - \half \sum_{q,\nu} i \omega_{q\nu} \Xi_{q\nu}^H \tilde{\mathcal{M}} \Xi_{q\nu} (b_{q,\nu}^\dagger b_{q,\nu} + 1/2),
\end{align}
where fluctuation terms of the form $b_{q \nu} b_{q \nu}$, $b_{q \nu}^\dagger b_{q \nu}^\dagger$ have been set to zero.
This expression can be compared to the usual expression for the quantum harmonic oscillator: $E=\hbar \omega (b^\dagger b + \half)$. Identifying the former expression with the latter results in the normalization:
\begin{align}
    \Xi_{q\nu}^H \tilde{\mathcal{M}} \Xi_{q\nu} = 2 i \hbar. \label{eq:M-normalization}
\end{align}
The total energy can be rewritten:
\begin{align}
    E = \sum_{q\nu} \hbar \omega_{q\nu} (b_{q,\nu}^\dagger b_{q,\nu} + 1/2).
\end{align}
With this, one can write the total energy for a single mode: $E_{q\nu}=-i \omega_{q\nu} \Xi_{q\nu}^H \mathcal{\tilde{M}} \Xi_{q\nu} / 2 = \hbar\omega_{q \nu}$.

The $\tilde{\mathcal{M}}$-normalization defined in Eq. \eqref{eq:M-normalization}, beyond establishing an orthonormalization of the eigenmodes, grants access to a natural decomposition of each mode into a phononic, magnonic, and coupled character based on their respective contributions to the total energy of the mode.
The $\tilde{\mathcal{M}}$-norm can be decomposed into
\begin{align}
    \Xi_{q\nu}^H \tilde{\mathcal{M}} \Xi_{q\nu}
    & = \biggl[
    \begin{pmatrix}
        \varepsilon^H & \mu^H
    \end{pmatrix}
    \begin{pmatrix}
        0 & -1 \\ 1 & 0
    \end{pmatrix}
    \begin{pmatrix}
        \varepsilon \\ \mu
    \end{pmatrix}\nonumber\\
    & \quad + \sigma^H \Gsstilde \sigma + \left(\varepsilon^H \Gustilde \sigma + \sigma^H \Gsutilde \varepsilon \right)\biggr].
\end{align}
From this decomposition, we can define the respective contributions to the total energy:
\begin{subequations}
\begin{align}
    & \eta_{q\nu}^{(\textrm{ph})} = \frac{1}{2i\hbar}
    \begin{pmatrix}
        \epsqv^H & \muqv^H
    \end{pmatrix}
    \begin{pmatrix}
        0 & -1 \\ 1 & 0
    \end{pmatrix}
    \begin{pmatrix}
        \epsqv \\ \muqv
    \end{pmatrix}, \label{eq:eta_ph}\\
    & \eta_{q\nu}^{(\textrm{mag})} = \frac{1}{2i\hbar} \sigmaqv^H \Gsstilde \sigmaqv, \label{eq:eta_mag}\\
    & \eta_{q\nu}^{(\textrm{cpld})} = \frac{1}{2i\hbar} \left(\epsqv^H \Gustilde \sigmaqv + \sigmaqv^H \Gsutilde \epsqv \right), \label{eq:eta_cpld}
\end{align}
    \label{eq:etas}
\end{subequations}
with the different contributions summing to one 
\begin{align}
    \eta_{q\nu}^{(\textrm{ph})} + \eta_{q\nu}^{(\textrm{mag})} + \eta_{q\nu}^{(\textrm{cpld})} = 1.
\end{align}
The energy for a single mode now becomes:
\begin{align}
    E_{q\nu} = \hbar \omega_{q\nu} [\eta_{q\nu}^{(\textrm{ph})} + \eta_{q\nu}^{(\textrm{mag})} + \eta_{q\nu}^{(\textrm{cpld})}].
\end{align}
This decomposition of hybrid modes into different characters (phononic, magnonic, coupled) is an improvement with respect to the current literature where no clear method surfaces, and the spin norm choice is often not explicit \cite{nomura19,Delugas23,Wang23,Bassant25}.
The phononic character can be further decomposed into atomic contributions
\begin{subequations}
\begin{align}
    \eta_{q\nu,\kappa}^{(\textrm{ph})} & = \frac{1}{2i\hbar} [\muqvk^H \epsqvk - \epsqvk^H \muqvk],
     \label{eq:atomic-mode-decomposition-a} \\
    \eta_{q\nu}^{(\textrm{ph})} & = \sum_\kappa \eta_{q\nu,\kappa}^{(\textrm{ph})}. \label{eq:atomic-mode-decomposition-b}
\end{align}
    \label{eq:atomic-mode-decomposition}
\end{subequations}
Such an atomic-resolved decomposition is generally not possible for the magnonic and the coupled character. However, it may still provide a good physical insight. Within the approximation of rigid atomic spins, the spin Berry curvature $\Gsstilde$ is block diagonal and the magnonic character can be decomposed exactly in atomic-resolved contributions \cite{niu99}.

Incidentally, $\eta_{q\nu}^\textrm{(mag)}$ corresponds also to the reduction $\Delta S_z = S_{0,z} - \expval{S_z}$ of the spin $z$-component with respect to the ground state due to the excitation of the mode $q\nu$ \cite{niu99}. In consequence, the spin angular momentum carried by a magnon is
\begin{align}
    l_{q\nu}^\textrm{(mag)} = \hbar \eta_{q\nu}^\textrm{(mag)} b_{q\nu}^\dagger b_{q\nu}.
\end{align}

Additionally to magnons, phonons can also carry angular momentum which in this case originates from the collective rotation of atoms around their equilibrium position.
The canonical angular momentum of a single atom can be written classically $l_{i} = u_{i} \times p_{i}$. Introducing the second quantization procedure from Section \ref{sub:second_quantization}, we obtain the expression for the phonon angular momentum
\begin{align}
    l_{q\nu}^\textrm{(ph)} = \Re\{\varepsilon_{q\nu,\kappa}^* \times \mu_{q\nu,\kappa}\} (b_{q\nu}^\dagger b_{q\nu} + \half).
\end{align}

\subsection{Nondimensionalization and reduction to pure phonons}

We now turn to the special case where phonons do not interact with magnons, but still incorporate effects of the time reversal symmetry breaking through the molecular Berry curvature $\Guu$. In this setting, the equations of motion Eq. \eqref{eq:EOM_quantized_short} become
\begin{align}
    -i\omegaqv
    \begin{pmatrix}
        0 & -1 \\
        1 &  0
    \end{pmatrix}
    \begin{pmatrix}
        \epsqv \\ \muqv
    \end{pmatrix}
    =
    \begin{pmatrix}
        \Duu & -\half \Guu \\
        \half \Guu & 1
    \end{pmatrix}
    \begin{pmatrix}
        \epsqv \\ \muqv
    \end{pmatrix}. \label{eq:EOM_TRSB_ph}
\end{align}
The normalization condition \eqref{eq:M-normalization} becomes
\begin{align}
    \muqv^H \epsqv - \epsqv^H \muqv = 2 ih. \label{eq:M-norm-ph}
\end{align}
Up to now, the eigenvectors $\epsqv,\ \muqv$ associated to the atomic displacements and momenta carried the dimensions of a mass-scaled length $LM^{1/2}$, and a mass-scaled length divided by a time $LM^{1/2}T^{-1}$. The units of the momenta incidentally correspond to that of an action divided by a mass-scaled length. To nondimensionalize the displacements we can define a characteristic mass-scaled length $\lqvzero$ such that $\epsqv = \lqvzero \epsqvtilde$.
Inserting this expression in the normalization condition \eqref{eq:M-norm-ph}, we see that in order to conserve the structure of the equation, we need to define $\muqv = \hbar \muqvtilde / \lqvzero$, which appropriately adimensionalize the momenta. We will further impose $\norm{\epsqvtilde}=1$. Multiplying the second line of Eq. \eqref{eq:EOM_TRSB_ph} by $\epsqv^H$ on the left, one obtains
\begin{align}
    -i\omegaqv \epsqv^H\epsqv = \half \epsqv^H\Guu\epsqv + \epsqv^H\muqv.
\end{align}
Subtracting the complex conjugate and rearranging the terms, one gets
\begin{align}
    \omegaqv \epsqv^H\epsqv - \frac{i}{2} \epsqv^H\Guu\epsqv = \frac{1}{2i}(\muqv^H\epsqv - \epsqv^H\muqv).
\end{align}
Using the normalization condition \eqref{eq:M-norm-ph} and replacing $\epsqv$ by its nondimensional form $\epsqvtilde$, we finally obtain the characteristic length
\begin{align}
    \lqvzero = \sqrt{\frac{\hbar}{\omegaqv - \frac{i}{2} \epsqvtilde^H\Guu\epsqvtilde}}. \label{eq:lqvzero}
\end{align}
In the case of time reversal symmetric phonons, it reduces to $\lqvzero = \sqrt{\hbar / \omegaqv}$. Furthermore, the definition of the atomic displacements and momenta Eq. \eqref{eq:Bogoliubov_displ} and \eqref{eq:Bogoliubov_momentum} becomes
\begin{align}
    u_{q \kappa} & = \sum_\nu \sqrt{\frac{\hbar}{2\omegaqv}} \left(\tilde \varepsilon_{q \nu, \kappa} b_{q \nu} + \tilde\varepsilon_{-q \nu, \kappa}^* b_{-q \nu}^\dagger \right),\\
    p_{q \kappa} & = \sum_\nu \sqrt{\frac{\hbar \omegaqv}{2}} \left(\tilde \mu_{q \nu, \kappa} b_{q \nu} + \tilde\mu_{-q \nu, \kappa}^* b_{-q \nu}^\dagger \right)
\end{align}
Both of these results agree with the textbook definition \cite{Bruesch82}.

Alternatively, instead of adimensionnalizing the momenta via $\muqv = \hbar \muqvtilde / \lqvzero$, one could have defined a characteristic frequency $\varpi_{q \nu}^0$ such that $\muqv = \muqvtilde \lqvzero \varpi_{q \nu}^0$. Inserting this expression in the normalization condition \eqref{eq:M-norm-ph}, one would have found $\varpi_{q \nu}^0 = \omegaqv - \frac{i}{2} \epsqvtilde^H\Guu\epsqvtilde$. This yields the same results as the procedure outlined above.

We now show that the decomposition of the mode energy as well as the atomic-resolved mode decomposition of the phononic character defined in Eq. \eqref{eq:atomic-mode-decomposition} is a generalization to the decomposition usually performed in the literature for pure time reversal symmetric phonons  \cite{ghosez99}.
Since we are dealing with purely phononic modes, we have $\eta_{q\nu}^{(\textrm{ph})}=1$. Restarting from Eq. \eqref{eq:atomic-mode-decomposition-a} and extracting $\muqv$ from the second line of Eq. \eqref{eq:EOM_TRSB_ph}, one obtains
\begin{align}
    \eta_{q\nu,\kappa}^{\textrm{(ph)}} = \epsqvk^H (\omegaqv - i\Guu/2) \epsqvk / \hbar.
\end{align}
Inserting the nondimensionalized $\epsqvtilde$ and using \eqref{eq:lqvzero}, the atomic-resolved mode decomposition reduces to
\begin{align}
    \eta_{q\nu,\kappa}^{\textrm{(ph)}} = \norm{\epsqvktilde}^2,
    \eta_{q\nu}^{(\textrm{ph})} = \sum_\kappa \eta_{q\nu,\kappa}^{\textrm{(ph)}} = 1.
\end{align}
This decomposition is valid for both time reversal symmetric and non time reversal symmetric phonons without phonon-magnon interaction. It matches the decomposition used in the literature for time reversal symmetric phonons \cite{ghosez99}.

\subsection{Link to Raman spin-lattice interaction}

In this paragraph we establish the link between the molecular Berry curvature and the Raman spin-lattice interaction, which was already discussed by \cite{qin12,saparov22}.
Since the molecular Berry curvature (MBC) is non-zero only in systems where the $\mathcal{T}$ symmetry is broken and that, in ferromagnetic materials, the symmetry breaking is due to its magnetization $\vec{M}$, it is natural to write the MBC as a function of $\vec{M}$: $G^{uu}=G^{uu}(\vec{M})$. Since the ground states with $+\vec{M}$ and $-\vec{M}$ magnetizations are linked by time reversal symmetry, we have $G^{uu}(\vec{M}) = - G^{uu}(-\vec{M})$. 
To first order in $\vec{M}$ the MBC can be expanded as
\begin{align}
    G^{uu}_{ij}(\vec{M}) = G^{uu}_{ij \delta} M_\delta.
\end{align}
Injecting this expression into the Hamiltonian of Eq. \eqref{eq:constrained_Ham} and isolating the term proportional to $p$ and $u$, one obtains: 
\begin{align}
    H_{pu} & = \half \sum_{ij \delta} p_i G^{uu}_{ij \delta} u_j M_\delta\\
    & = \half \sum_{ll'} \sum_{\kappa \alpha, \kappa' \beta, \delta} p_{l \kappa \alpha} G^{uu}_{l \kappa \alpha, l' \kappa' \beta, \delta} u_{l' \kappa' \beta} M_\delta,
\end{align}
where $\alpha,\beta,\delta$ correspond to cartesian directions.
This term is non-local in general. Restricting ourselves to the on-site contribution ($l \kappa = l' \kappa'$), we can make use of the skew-symmetry of the MBC to write $G^{uu}_{l \kappa \alpha, l \kappa \beta, \delta}=\varepsilon_{\alpha \beta \gamma} g^{uu}_{l \kappa, \gamma, \delta}$. This effectively transforms the matrix-vector product as a vector cross-product. Developing the above expression gives:
\begin{align}
    H_{pu}^\textrm{loc} & = \half \sum_{l \kappa, \alpha \beta \gamma, \delta} p_{l \kappa \alpha} ( \varepsilon_{\alpha \beta \gamma} g^{uu}_{l \kappa, \gamma, \delta} ) u_{l \kappa \beta} M_\delta\\
    & = \half \sum_{l \kappa, \delta} \boldsymbol{p}_{l \kappa} \cdot ( \boldsymbol{g}^{uu}_{l \kappa, \delta} \times \boldsymbol{u}_{l \kappa} ) M_\delta\\
    & = \half \sum_{l \kappa, \delta} ( \boldsymbol{u}_{l \kappa} \times \boldsymbol{p}_{l \kappa} ) \cdot \boldsymbol{g}^{uu}_{l \kappa, \delta} M_\delta\\
    & = \half \sum_{l \kappa, \gamma \delta} L_{l \kappa \gamma} g^{uu}_{l \kappa, \gamma \delta} M_\delta.
\end{align}
Further assuming an isotropic $g^{uu}$, this can be simplified into:
\begin{align}
    H_{pu}^\textrm{loc} = \half \sum_{l \kappa} g^{uu}_{l \kappa} \boldsymbol{L}_{l \kappa} \cdot \boldsymbol{M},
\end{align}
which is the expression of the Raman spin-lattice interaction under the mean field approximation~\cite{sheng06}.

\begin{figure}
     \centering
     \includegraphics[height=0.4\textheight]{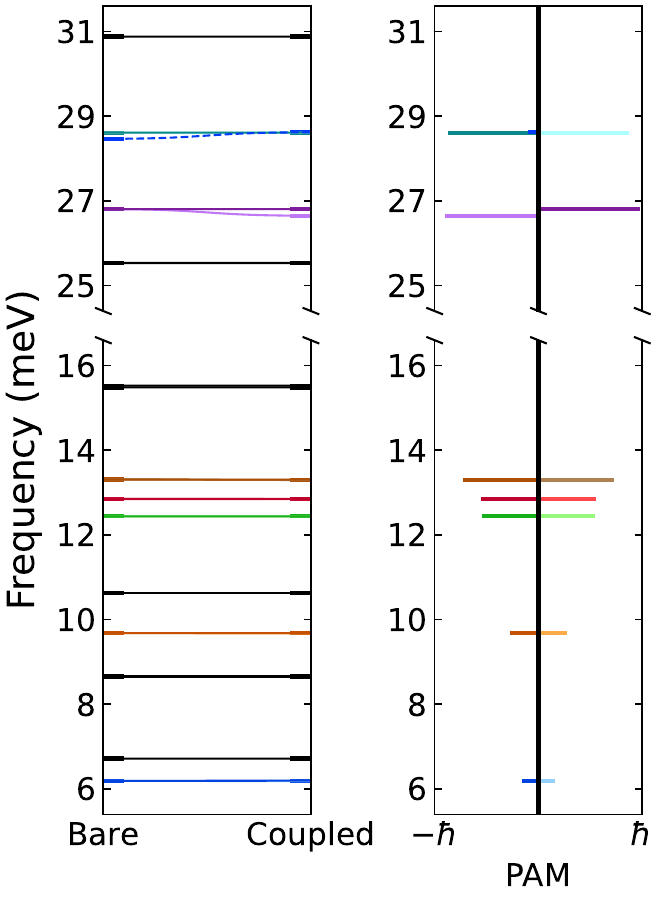}
     \caption{Frequency splitting and phonon angular momentum for CrI3 at the $\Gamma$ point. (a) Bare and coupled frequencies for the phonons and the optical magnon. Non-degenerate phonon $A$-modes are in black, doubly-degenerate $E$-modes in light/dark color pairs, and the optical magnon in dashed dark blue around $28.47$ meV. Non-degenerate phonon modes undergo no shift in frequency. Degenerate phonon modes show a slight splitting at the exception of the pair at $26.81$ meV in violet/purple where the $E_u^-$ mode shifts downward significantly in response to the coupling with the optical magnon. The optical magnon shifts upward by a similar amplitude. (b) Phonon angular momentum for each mode in the coupled regime. The non-degenerate phonon modes do not acquire any PAM. Degenerate phonons acquire significant PAM through the splitting in frequency. For each pair, the PAM has almost the same amplitude but with opposite signs. For the violet/purple pair, the PAM has still opposite signs but no longer the same amplitudes. This is a consequence of the phonon-magnon hybridization between the $E_u^-$ and the optical magnon, where the optical magnon gains phononic character and steals PAM at the same time.}
     \label{fig:CrI3}
\end{figure}

\begin{figure}
     \centering
     \includegraphics[height=0.4\textheight]{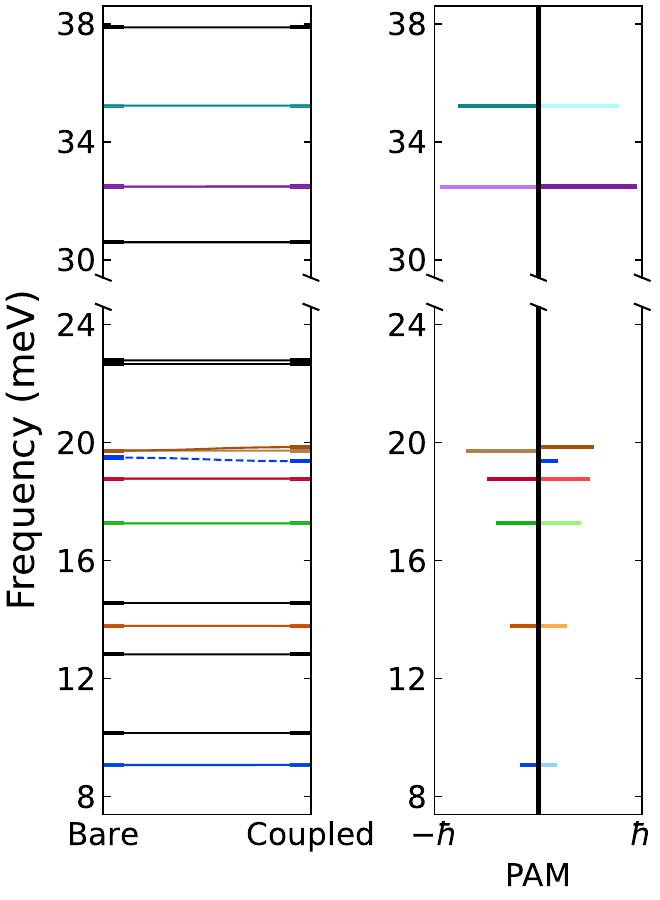}
     \caption{Frequency splitting and phonon angular momentum for CrBr$_3$ at the $\Gamma$ point. (a) Bare and coupled frequencies for the phonons and the optical magnon. Non-degenerate phonon $A$-modes are in black, doubly-degenerate $E$-modes in light/dark color pairs, and the optical magnon in dashed dark blue around $19.49$ meV. (b) Phonon angular momentum for each mode in the coupled regime. The non-degenerate phonon modes do not acquire any PAM. Degenerate phonons acquire significant PAM through the splitting in frequency.}
     \label{fig:CrBr3}
\end{figure}

\section{Results}\label{sec:results}

We performed \textit{Density Functional Theory} calculations on the ferromagnetic semiconducting Van der Waals materials CrI$_3$ and CrBr$_3$ \cite{mcguire15,mcguire17}. We used the \textsc{Abinit} software \cite{gonze02,verstraete25} with the Perdew and Wang correlation functional \cite{perdew92a}. 
We used a planewave basis set with an energy cutoff of $40$ Ha and a $6\times6\times6$ $k$-point grid. The lattice parameters and atomic positions have been relaxed to achieve forces on atoms under $10^{-6}$ Ha/Bohr. We used constrained magnetic moment DFPT to perform calculations at fixed spin and obtain derivatives with respect to atomic spins \cite{Royo26}. All calculations were performed at $q=\Gamma$.

CrI$_3$ and CrBr$_3$ belong to the space group $R\bar 3$ and possess the spatial inversion symmetry (however they do not have time reversal symmetry due to their magnetization). At the $\Gamma$ point, the phonon modes can be classified according to their irreducible representations (irreps): 4 $A_g$ and 4 $A_u$ modes, plus 4 $E_g$ and 4 $E_u$ pairs of doubly-degenerate modes. One $A_u$ mode and one pair of $E_u$ modes are acoustic phonons, while the rest are optical phonons \cite{bonini23,ren24a}. Since CrI$_3$ and CrBr$_3$ have two magnetic Cr ions per unit cell, they have one acoustic (AM) and one optical (OM) magnon mode each.

Both CrI$_3$ and CrBr$_3$ present a C$_3$ rotational axis along the $z$ direction. Therefore, at $q$~points invariant under the C$_3$ symmetry, it is useful to introduce the crystal angular momentum (or pseudoangular momentum) associated to the rotational axis. It has been shown that crystal angular momentum (CAM) must be conserved during hole-phonon, light-phonon and phonon-phonon scattering processes~\cite{zhu18,Ishito22,Minakova26,juraschek25}. Applying the definition from Ref.~\cite{Zhang15} to complete phonon-magnon eigenvectors, the crystal angular momentum $\chi_{q\nu}^{(n)}$ associated to a C$_n$ axis is given by
\begin{align}
    C_n \Xi_{q\nu} = e^{i(2\pi/3)\chi_{q\nu}^{(n)}} \Xi_{q\nu}.
\end{align}
Using the $\tilde{\mathcal{M}}$-norm defined in Eq.~\eqref{eq:M-normalization}, it can be reframed as
\begin{align}
    \frac{\Xi_{q\nu}^H \tilde{\mathcal{M}} C_n \Xi_{q\nu}}{\Xi_{q\nu}^H \tilde{\mathcal{M}} \Xi_{q\nu}} = e^{i(2\pi/3)\chi_{q\nu}^{(n)}}.
\end{align}
From this last equation, the crystal angular momentum $\chi_{q\nu}^{(n)}$ can be easily extracted. For a C$_3$ axis, the CAM can take the three value $0$, $+1$ and $-1$. In scattering processes, CAM is conserved modulo $n$, or modulo $3$ in this case~\cite{Ishito22,Minakova26}.

The bare phonon and magnon frequencies can be obtained by solving the uncoupled equations of motion Eq. \eqref{eq:bare_phonon} and \eqref{eq:bare_magnon}.
In this setting, the angular momentum of phonons is zero and the angular momentum of magnons is equal to $\hbar$.
The obtained bare phonon frequencies are in correct agreement with Raman scattering experiments and previous DFT calculations \cite{djurdjic-mijin18,larson18,Kozlenko21,Kipczak24}.
A comparison table can be found in Appendix \ref{sec:comp_exp_values}.
The optical magnon frequencies predicted here at $28$ and $19$ meV respectively for CrI$_3$ and CrBr$_3$ deviate significantly from the experimental values of $19$ and $13$ meV \cite{Chen18,Nikitin22}.
This is a known limitation of DFT when using an LDA functional \cite{Ke21} and is consistent with previous DFT calculations \cite{Ke21,Gorni23,Royo26}. 

The coupled phonon-magnon frequencies are obtained by solving the full equations of motion \eqref{eq:EOM_quantized_short}. The bare and coupled frequencies are reported in Fig.~\ref{fig:CrI3}(a) for CrI$_3$ and Fig.~\ref{fig:CrBr3}(a) for CrBr$_3$. From the molecular Berry curvature and the coupling with the magnons, the phonons are able to obtain non-zero angular momentum, which is reported in Figs.~\ref{fig:CrI3}(b),\ref{fig:CrBr3}(b).
Up to numerical accuracy, non-degenerate phonons (with $A_g$ or $A_u$ irreps) undergo no shift in frequency or hybridization with the magnons. Doubly-degenerate phonons (with $E_g$ or $E_u$ irreps) are split in frequency by the inclusion of the molecular Berry curvature and the coupling with magnons. Each pair is split into a left and a right circularly polarized mode. The largest frequency splitting occurs in the $E_u$ phonon mode pair the closest in frequency to the optical magnon. This is in agreement with the conclusions of Ref. \cite{ren24a}. The values of the frequency splittings and the phonon angular momentum are summarized in table \ref{tab:CrI3_Emodes} and \ref{tab:CrBr3_Emodes}.

\begin{table}[t]
\centering
\caption{Bare frequencies, induced splitting/shift, phonon angular momentum and crystal angular momentum for the $E$ phonon modes and the magnon modes for CrI$_3$.}
\label{tab:CrI3_Emodes}
\begin{tabular}{ccccr}
  Irrep & $E_0$(meV) & $\Delta E$(meV) & $L_z$ ($\hbar$) & $\chi_{q\nu}^{(3)}$ \\ \hline
  $E_g$ &  6.19 & 0.0079 &  .1401 &  1 \\
        &       &        & -.1412 & -1 \\
        & 12.44 & 0.0001 &  .5277 & -1 \\
        &       &        & -.5272 &  1 \\
        & 12.85 & 0.0045 &  .5332 & -1 \\
        &       &        & -.5327 &  1 \\
        & 28.61 & 0.0047 &  .8539 &  1 \\
        &       &        & -.8539 & -1 \\ \hline
  $E_u$ &  9.68 & 0.0022 &  .2550 & -1 \\
        &       &        & -.2548 &  1 \\
        & 13.31 & 0.0071 &  .7037 & -1 \\
        &       &        & -.7051 &  1 \\
        & 26.81 & 0.1582 & -.8824 & -1 \\
        &       &        &  .9560 &  1 \\ \hline
  AM    &  0.80 & 0.0105 & $2.584 \times 10^{-5}$ & -1 \\ \hline
  OM    & 28.47 & 0.1566 & -.0762 & -1 \\
\end{tabular}
\end{table}

\begin{table}[t]
\centering
\caption{Bare frequencies, induced splitting/shift, phonon angular momentum and crystal angular momentum for the $E$ phonon modes and the magnon modes for CrBr$_3$.}
\label{tab:CrBr3_Emodes}
\begin{tabular}{ccccr}
  Irrep & $E_0$ (meV) & $\Delta E$ (meV) & $L_z$ ($\hbar$) & $\chi_{q\nu}^{(3)}$ \\ \hline
  $E_g$ &  9.07 & 0.0004 &  .1555 &  1 \\
        &       &        & -.1557 & -1 \\
        & 17.26 & 0.0002 &  .3884 & -1 \\
        &       &        & -.3883 &  1 \\
        & 18.78 & 0.0008 &  .4761 &  1 \\
        &       &        & -.4760 & -1 \\
        & 35.23 & 0.0005 &  .7567 &  1 \\
        &       &        & -.7566 & -1 \\ \hline
  $E_u$ & 13.79 & 0.0020 &  .2531 & -1 \\
        &       &        & -.2530 &  1 \\
        & 19.73 & 0.1212 & -.6773 &  1 \\
        &       &        &  .5138 & -1 \\
        & 32.49 & 0.0038 & -.9301 & -1 \\
        &       &        &  .9303 &  1 \\ \hline
  AM    &  0.22 &-0.0002 & $-6.911 \times 10^{-6}$ & -1 \\ \hline
  OM    & 19.49 &-0.1233 &  .1632 & -1 \\
\end{tabular}
\end{table}

\begin{table}[t]
\centering
\caption{Bare frequencies, induced shifts, and decomposition into phonon, magnon and coupled character for selected $E_u$ phonon modes and for the optical magnon in CrI$_3$. The different characters are computed from Eq. \eqref{eq:etas} and are presented in \%.}
\label{tab:CrI3_phmag_charac}
\begin{tabular}{|c|c|c|c|c|c|}
  \hline
  Irrep   & $E_0$ (meV) & $\Delta E$ (meV) & $\eta^{(\textrm{ph})}$ & $\eta^{(\textrm{mag})}$ & $\eta^{(\textrm{cpld})}$ \\ \hline
  $E_u^-$ & 26.81 & -0.1586 &  92.239 &  7.772 & -0.011 \\ \hline
  $E_u^+$ &       &  0.0003 & 100.010 & -0.010 &  0.000 \\ \hline
  OM      & 28.47 &  0.1566 &   7.828 & 92.161 &  0.011 \\ \hline
\end{tabular}
\end{table}

\begin{table}[t]
\centering
\caption{Bare frequencies, induced shifts, and decomposition into phonon, magnon and coupled character for selected $E_u$ phonon modes and for the optical magnon in CrBr$_3$. The different characters are computed from Eq. \eqref{eq:etas} and are presented in \%.}
\label{tab:CrBr3_phmag_charac}
\begin{tabular}{|c|c|c|c|c|c|}
  \hline
  Irrep   & $E_0$ (meV) & $\Delta E$ (meV) & $\eta^{(\textrm{ph})}$ & $\eta^{(\textrm{mag})}$ & $\eta^{(\textrm{cpld})}$ \\ \hline
  $E_u^-$ & 19.73 &  0.1200 &  74.984 & 25.007 &  0.009 \\ \hline
  $E_u^+$ &       & -0.0012 & 100.003 & -0.003 &  0.000 \\ \hline
  OM      & 19.49 & -0.1233 &  25.047 & 74.962 & -0.009 \\ \hline
\end{tabular}
\end{table}

Using Eq. \eqref{eq:eta_ph}, \eqref{eq:eta_mag} and \eqref{eq:eta_cpld}, we can compute the phononic, magnonic and hybrid character of each mode. In CrI$_3$, all phonons remain pure to within $\pm0.1\%$, except one of the modes in the $E_u$ pair at $26.81$ meV (see Table \ref{tab:CrI3_phmag_charac}). The phonon mode hybridizes and obtains $\sim8\%$ magnonic character, and the optical magnon incorporates around the same amount of phononic character.
Similarly in CrBr$_3$, all phonon modes remain phonon-like within $\pm0.1\%$, except for one of the $E_u$ modes at $19.73$ meV, which gains $\sim25\%$ magnonic character. This large fraction is exchanged with the optical magnon which itself recovers $25\%$ phononic character (see Table \ref{tab:CrBr3_phmag_charac}).
The fact that only one phonon from the $E_u$ pair hybridizes with the optical magnon is a cornerstone of chirality selective phonon-magnon coupling. Phonons and magnons with the same CAM numbers (or handedness) will interact, whereas phonons and magnons with opposite CAM numbers (or opposite handedness) interact much more weakly \cite{Gurevich96,Weissenhofer25,cui23,ren24a}.

From Tables \ref{tab:CrI3_Emodes} and \ref{tab:CrBr3_Emodes}, we see that the angular momentum within one $E$ pair of phonon modes almost adds up to zero and shows only a small imbalance. In contrast, the $E_u$ pair that couples to the optical magnon shows a much larger imbalance. This is explained by the fact that the $E_u$ phonon that gains some magnonic character loses an equal amount of phononic character to the optical magnon, and with it some of its phonon angular momentum.
As a consequence of gaining phononic character, the optical magnon also gains phonon angular momentum, which can be seen in Tables \ref{tab:CrI3_Emodes} and \ref{tab:CrBr3_Emodes}.
During this transfer with the $E_u$ mode, PAM is conserved within $\sim 10^{-3}\,\hbar$. When accounting for transfer between the optical magnon and all phonons, PAM is conserved within $\sim 10^{-4}\,\hbar$. The conservation of PAM is no longer an absolute rule due to a local metric modification by the mixed spin phonon Berry curvature.

\section{Conclusion}\label{sec:conclusion}

Building upon the Lagrangian formulation of coupled spin-phonon dynamics, we derive a constrained Hamiltonian theory.
This allows us to derive more symmetric and numerically more stable equations of motion.
More significantly, this framework allows direct access to the total energy. By decomposing the total energy into phononic, magnonic and coupled contributions, we are able to quantify the degree of hybridization of phonon-magnon modes. We find phonon-magnon pairs with up to $8\%$ and $25\%$ hybridization for bulk CrI$_3$ and CrBr$_3$ respectively.
We particularize our findings to pure non-time reversal symmetric phonons which include effects of the molecular Berry curvature but exclude the coupling with magnons.
We show that the quantification of the hybridization through the decomposition of the total energy is a generalization to the decomposition of the norm of the displacement eigenvectors that is usually employed in the literature.
We also derive an orthonormalization condition for the hybrid phonon-magnon eigenvectors. This step had been overlooked up to now in the literature, but is crucial for the accurate computation of quantities relying on the mode eigenvectors.
An example of this is the phonon angular momentum that we compute in both bulk CrI$_3$ and CrBr$_3$. We find that, through hybridization with circularly polarized phonons, magnons are able to steal part of the phonon mechanical angular momentum. This stealing mechanism was not accounted for in current literature using naive orthonormalization schemes.
In the future, we aim to extend the results obtained here to the whole Brillouin zone. This gives access to the full characterization of anti-crossing points between phonon and magnon bands. It allows the computation of band topology and topological transport as well \cite{Thingstad19}.

An important characteristic of the current theory is that we do not employ the rigid spin approximation when incorporating adiabatic corrections. One could easily relax the atomic spin picture used here, as initially performed by Niu and Kleinman \cite{niu98}, to study materials with weakly localized $4d$ or $5d$ orbitals where the rigid atomic spin picture does not hold anymore \cite{Lin25}.
The framework developed here can also be generalized and applied to other systems of composite hybrid particles such as magnon-plasmon \cite{Costa23,Hirosawa26} and magnon-exciton hybrids\cite{Adak26}.

We foresee that the theoretical tools developed in this paper will also prove useful to study the transport of phonon angular momentum in magnetic systems. These tools, and notably the stealing mechanism outlined in this paper, will enable the study of conversion of angular momentum between magnetic degrees of freedom and the underlying ionic lattice, for example in ultrafast demagnetization processes \cite{tauchert22}. The theory developed so far only deals with equilibrium states, and one would need to include scattering through higher order force constants, and coupling with external driving sources (light, electrical or spin currents, temperature gradient).

\paragraph{Code and data availability}
The code for obtaining the results presented in this paper is available at \url{https://www.github.com/MaxMignolet/Hyph-m}. The code repository also contains the stiffness matrices and Berry curvatures to compute the hybrid phonon-magnon modes in CrI$_3$ and CrBr$_3$. The Abinit input files employed to obtain these are also present.



\paragraph{Funding information}
This work was supported by the Fonds National de la Recherche Scientifique (FRS-FNRS Belgium) through the Excellence of Science (EOS) program (Grant No. 40007563 CONNECT) and by the Fédération Wallonie Bruxelles through the ARC project DREAMS (G.A. 21/25-11).
M.M. is an Aspirant PhD candidate of the Fonds de la Recherche Scientifique - FNRS, and received a Scientific Stay grant from the FNRS under number 2024/V 5/6/026 - JG/DeM - 2615.
M.J.V. acknowledges funding by the Dutch Gravitation program “Materials for the Quantum Age” (QuMat, reg number 024.005.006), financed by the Dutch Ministry of Education, Culture and Science (OCW).
M.R and M.S. acknowledge support from the Spanish MCIN/AEI/10.13039/501100011033 through grants
PID2019-107338RB-C61,
PID2022-139776NBC65,
PRX22/00390, and
PID2023-152710NB-I00
and a Severo Ochoa Excellence award to ICMAB, Grant CEX2023-001263-S,
as well as by the Generalitat de Catalunya through Grant No. 2021 SGR 01519.
Computational resources have been provided
by PRACE (Optospin Project Id. 2020225411) on Discoverer at Sofiatech Bulgaria;
by EuroHPC (Extreme Grant No. EHPCEXT-2023E02-050) on Marenostrum5 at Barcelona Supercomputing Center (BSC), Spain;
on Lucia, the Tier-1 supercomputer of the Walloon Region, infrastructure funded by the Walloon Region under the grant agreement n°1910247;
by the Dutch NWO for access to the LUMI supercomputer, owned by
EuroHPC-JU and hosted by CSC (FI) through the ‘National Computer Facilities’ call (SURF reference EINF-16505).

\begin{appendix}
\numberwithin{equation}{section}

\section{Comparison of bare frequencies with experimental values} \label{sec:comp_exp_values}

In this appendix we compare the ab initio phonon frequencies obtained in the main text with those obtained experimentally in the literature.
A first comparison of the computed bare phonon frequencies in CrI$_3$ against values obtained in Ref. \cite{djurdjic-mijin18} is shown in table \ref{tab:CrI3_ph_comp}.
We use the bare phonon frequencies as to simplify the analysis since for the majority of the modes there is little change between the bare and the coupled frequencies. We also compare with other first-principles works. Overall, the obtained frequencies are in good agreement, at the exception of the $A_g^2$ mode which underestimate the measured frequency by about $20\%$. This can be explained by the fact that we employ an LDA exchange-correlation functional and that we did not use a Van der Waals correction. This is in contrast to Ref. \cite{djurdjic-mijin18,larson18} where they use DFT-D Van der Waals corrections as well as GGA functionals. 

In table \ref{tab:CrBr3_ph_comp}, we compare the phonon frequencies obtained in the main text for CrBr$_3$ with experimental data from Ref. \cite{Kozlenko21} and DFT calculations from Ref. \cite{Kipczak24}. The obtained frequencies matches well with the experiment although there is a noticeable deviation from the DFT calculations from Ref. \cite{Kipczak24} at higher frequencies.

\begin{table}[t]
\centering
\caption{Comparison table for the bare phonon frequencies obtained for CrI$_3$ by solving Eq. \eqref{eq:bare_phonon} with experimental and numerical values from \cite{djurdjic-mijin18,larson18}. All values are in cm$^{-1}$.}
\label{tab:CrI3_ph_comp}
\begin{tabular}{|c|c|c|c|c|}
  \hline
  Irrep   & w$_0$ & Exp. \cite{djurdjic-mijin18} & DFT \cite{djurdjic-mijin18} & DFT \cite{larson18} \\ \hline
  $E_g^1$ &  49.9 &  54.1 &  59.7 &  53   \\ \hline 
  $A_u^1$ &  54.2 &       &       &  64   \\ \hline
  $A_g^1$ &  69.8 &  73.3 &  89.6 &  79   \\ \hline
  $E_u^1$ &  78.1 &       &       &  83   \\ \hline
  $A_g^2$ &  85.8 & 108.3 &  98.8 &  88   \\ \hline
  $E_g^2$ & 100.3 & 102.3 &  99.8 &  98   \\ \hline
  $E_g^3$ & 103.6 & 106.2 & 112.2 & 102   \\ \hline
  $E_u^2$ & 107.4 &       &       & 107   \\ \hline
  $A_g^3$ & 124.9 & 128.1 & 131.1 & 125   \\ \hline
  $A_u^2$ & 125.2 &       &       & 123   \\ \hline
  $A_g^4$ & 206.0 & -     & 195.2 & 195   \\ \hline
  $E_u^3$ & 216.2 &       &       & 206   \\ \hline
  $E_g^4$ & 230.8 & 236.6 & 234.4 & 225   \\ \hline
  $A_u^3$ & 249.1 &       &       & 240   \\ \hline
\end{tabular}
\end{table}

\begin{table}[t]
\centering
\caption{Comparison table for the bare phonon frequencies obtained for CrBr$_3$ by solving Eq. \eqref{eq:bare_phonon} with experimental and numerical values from \cite{Kozlenko21,Kipczak24}. All values are in cm$^{-1}$.}
\label{tab:CrBr3_ph_comp}
\begin{tabular}{|c|c|c|c|}
  \hline
  Irrep   & w$_0$ & Exp. \cite{Kozlenko21} & DFT \cite{Kipczak24} \\ \hline
  $E_g^1$ &  73.2 &  74.7 &  71.3 \\ \hline 
  $A_u^1$ &  82.0 &       &  83.8 \\ \hline
  $A_g^1$ & 103.4 & 107.0 & 105.5 \\ \hline
  $E_u^1$ & 111.2 &       & 111.0 \\ \hline
  $A_g^2$ & 117.4 &       & 116.2 \\ \hline
  $E_g^2$ & 139.2 & 143.4 & 137.3 \\ \hline
  $E_g^3$ & 151.5 & 153.3 & 141.8 \\ \hline
  $E_u^2$ & 159.1 &       & 149.9 \\ \hline
  $A_u^2$ & 182.9 &       & 174.9 \\ \hline
  $A_g^3$ & 183.8 & 184.7 & 177.3 \\ \hline
  $A_g^4$ & 246.8 &       & 222.7 \\ \hline
  $E_u^3$ & 262.0 &       & 240.2 \\ \hline
  $E_g^4$ & 284.2 & 285.0 & 266.3 \\ \hline
  $A_u^3$ & 305.6 &       & 290.1 \\ \hline
\end{tabular}
\end{table}

\section{Additional data}

In this appendix, we put the extended tables \ref{tab:CrI3_Emodes_full} and \ref{tab:CrBr3_Emodes_full} regrouping the induced splitting, phonon angular momentum and hybridization characters for the phonon $E$ modes and magnon modes for both bulk CrI$_3$ and CrBr$_3$.

\begin{table}[t]
\centering
\caption{Bare frequencies, induced splitting/shift, phonon angular momentum and phononic, magnonic and coupled characters for the $E$ phonon modes and the magnon modes for CrI$_3$.}
\label{tab:CrI3_Emodes_full}
\begin{tabular}{|c|c|c|c|c|c|c|}
  \hline
  Irrep & $E_0$ & $\Delta E$ & $L_z$ ($\hbar$) & $\eta^{(\textrm{ph})}$ & $\eta^{(\textrm{mag})}$ & $\eta^{(\textrm{cpld})}$ \\ \hline
  $E_g$ &  6.19 & 0.0087 &  .1399    & 100.043 & -0.041 & -0.002 \\ \hline
  $E_g$ &  6.19 & 0.0079 &  .1401    & 100.043 & -0.043 & -0.001 \\ \hline
        &       &        & -.1412    &  99.925 &  0.076 & -0.001 \\ \hline
        & 12.44 & 0.0001 &  .5277    & 100.000 &  0.000 &  0.000 \\ \hline
        &       &        & -.5272    & 100.000 & -0.000 &  0.000 \\ \hline
        & 12.85 & 0.0045 &  .5332    & 100.012 & -0.012 &  0.000 \\ \hline
        &       &        & -.5327    &  99.986 &  0.014 &  0.000 \\ \hline
        & 28.61 & 0.0047 &  .8539    & 100.000 & -0.000 & -0.000 \\ \hline
        &       &        & -.8539    & 100.000 &  0.000 & -0.000 \\ \hline
  $E_u$ &  9.68 & 0.0022 &  .2550    &  99.987 &  0.013 &  0.000 \\ \hline
        &       &        & -.2548    & 100.003 & -0.003 & -0.000 \\ \hline
        & 13.31 & 0.0071 &  .7037    &  99.923 &  0.079 & -0.002 \\ \hline
        &       &        & -.7051    & 100.011 & -0.011 &  0.001 \\ \hline
        & 26.81 & 0.1582 & -.8824    &  92.239 &  7.772 & -0.011 \\ \hline
        &       &        &  .9560    & 100.010 & -0.010 &  0.000 \\ \hline
  AM    &  0.80 & 0.0105 & 2.584E-05 &   0.035 & 99.964 &  0.001 \\ \hline
  OM    & 28.47 & 0.1566 & -.0762    &   7.828 & 92.161 &  0.011 \\ \hline
\end{tabular}
\end{table}

\begin{table}[t]
\centering
\caption{Bare frequencies, induced splitting/shift, phonon angular momentum and phononic, magnonic and coupled characters for the $E$ phonon modes and the magnon modes for CrBr$_3$.}
\label{tab:CrBr3_Emodes_full}
\begin{tabular}{|c|c|c|c|c|c|c|}
  \hline
  Irrep & $E_0$ & $\Delta E$ & $L_z$ ($\hbar$) & $\eta^{(\textrm{ph})}$ & $\eta^{(\textrm{mag})}$ & $\eta^{(\textrm{cpld})}$ \\ \hline
  $E_g$ &  9.07 & 0.0004 &  .1555 & 100.001 & -0.000 & -0.000 \\ \hline
        &       &        & -.1557 & 100.001 & -0.000 & -0.000 \\ \hline
        & 17.26 & 0.0002 &  .3884 & 100.000 &  0.000 &  0.000 \\ \hline
        &       &        & -.3883 & 100.000 & -0.000 &  0.000 \\ \hline
        & 18.78 & 0.0008 &  .4761 & 100.000 & -0.000 &  0.000 \\ \hline
        &       &        & -.4760 & 100.000 &  0.000 &  0.000 \\ \hline
        & 35.23 & 0.0005 &  .7567 & 100.000 & -0.000 &  0.000 \\ \hline
        &       &        & -.7566 & 100.000 &  0.000 &  0.000 \\ \hline
  $E_u$ & 13.79 & 0.0020 &  .2531 &  99.983 &  0.017 &  0.000 \\ \hline
        &       &        & -.2530 & 100.000 & -0.000 & -0.000 \\ \hline
        & 19.73 & 0.1212 & -.6773 & 100.003 & -0.003 &  0.000 \\ \hline
        &       &        &  .5138 &  74.984 & 25.007 &  0.009 \\ \hline
        & 32.49 & 0.0038 & -.9301 &  99.981 &  0.019 &  0.000 \\ \hline
        &       &        &  .9303 & 100.001 & -0.001 &  0.000 \\ \hline
  AM    &  0.22 &-0.0002 & -0.6911E-06 &   0.000 & 100.000 &  0.000 \\ \hline
  OM    & 19.49 &-0.1233 &  .1632 &  25.048 & 74.962 & -0.009 \\ \hline
\end{tabular}
\end{table}
\end{appendix}





\bibliography{bibliography.bib}


\end{document}